\documentclass[twocolumn, prl, superscriptaddress]{revtex4-1}
\usepackage{graphicx}
\usepackage{dcolumn}
\usepackage{bm}
\usepackage{hyperref}
\usepackage[colorinlistoftodos]{todonotes}
\usepackage{longtable}
\presetkeys{todonotes}{inline,backgroundcolor=yellow}{}
\usepackage[normalem]{ulem}

\begin{document}
\title{Enhanced sensitivity of the electron electric dipole moment from YbOH: The role of theory}
\author{V. S. Prasannaa}
\affiliation{Physical Research Laboratory, Atomic, Molecular and Optical Physics Division, Navrangpura, Ahmedabad-380009, India}

\author{N. Shitara}
\affiliation{Department of Physics, Tokyo Institute of Technology,
2-1-I-H86 Ookayama, Meguro-ku, Tokyo 152-8550, Japan}

\author{A. Sakurai}
\affiliation{Department of Physics, Tokyo Institute of Technology,
2-1-I-H86 Ookayama, Meguro-ku, Tokyo 152-8550, Japan}

\author{M. Abe}
\affiliation{Tokyo Metropolitan University, 1-1, Minami-Osawa, Hachioji-city, Tokyo 192-0397, Japan}

\author{B. P. Das}
\affiliation{Department of Physics, Tokyo Institute of Technology,
2-1-I-H86 Ookayama, Meguro-ku, Tokyo 152-8550, Japan}

\date{\today}

\begin{abstract}
The prospect of laser cooling of polyatomic molecules has opened a new avenue in the search for the electric dipole moment of the electron (eEDM). An upper bound on the eEDM would probe new physics arising from beyond the Standard Model of elementary particles. In this work, we report the first theoretical results for the effective electric field experienced by the electron in YbOH, and its molecular electric dipole moment, using a relativistic coupled cluster theory. We compare these two properties of YbOH with YbF, which also has a singly unoccupied orbital on the Yb ion. We also present the results of the effective electric field for different bond angles, which sheds light on the sensitivity that can be expected from an eEDM experiment with YbOH.  
\end{abstract}

\maketitle


The electric dipole moment of the electron (eEDM) could arise due to simultaneous violations of parity (P) and time reversal (T) symmetries~\cite{Landau,Ballentine,Sandars,Krip}. Therefore, if observed, the eEDM would provide a direct proof of $T$-violation. In spite of many ingenious experiments for over five decades, this property has not yet been observed. Currently, heavy polar diatomic molecules provide the best upper bounds on the eEDM~\cite{DeMilleReview,ChuppReview}, with the best result coming from ThO~\cite{Baron2014,Baron2017,Baron2018}, followed by HfF$^+$~\cite{Cairncross2017} and YbF~\cite{Hudson2011}. 
The Standard Model backgrounds for the eEDM are ten orders of magnitude below the current experimental limit~\cite{Pospelov}. However, the eEDM values predicted by theories beyond the Standard Model are much larger~\cite{SUSY}, and most of them are well within the bounds set by the eEDM experiments to date. These bounds, therefore, constrain theories beyond the Standard Model~\cite{Fukuyama}, as well as offer insights into the baryon asymmetry in the universe~\cite{Fuyuto}. The importance of this approach to furthering our understanding of new physics stems from the fact that one can probe PeV energy scales without using high energy accelerators~\cite{Nath}, but instead using molecular table-top experiments that measure quantities like the eEDM to very high levels of precision. \\

A measurement of the shift in energy of a molecule ($\Delta E$) in some state due to an electron's EDM ($d_e$), in combination with a theoretically determined effective electric field, $\mathcal{E}_\mathrm{eff}$ (related by the expression $\Delta E = - d_e \mathcal{E}_\mathrm{eff}$), yields an upper bound to the eEDM. However, the choice of a candidate molecule for an experiment depends on various considerations, which include a fairly large $\mathcal{E}_\mathrm{eff}$, and a reasonable molecular electric dipole moment (\emph{not} P and T violating, and denoted in this manuscript by $d_M$). $d_M$ plays a key role in deciding the extent to which one can polarize a molecule in a lab frame (quantified by the polarization factor, $\eta$). Specifically, $\eta$ is proportional to $d_ME/\Delta$, where E is the applied electric field, and $\Delta$ is the energy difference between opposite parity states in the molecule. \\

The choice of a candidate molecule for an eEDM experiment  also relies upon experimental factors such as the number of molecules (N) that one can produce, the total integration time (T), coherence times of the molecule ($\tau$), as well as robustness to systematic errors. Of these factors, $\mathcal{E}_\mathrm{eff}$ plays a special role, and this can be understood from the expression for the figure of merit for the statistical sensitivity in an eEDM experiment \\

\begin{eqnarray}
\mathcal{F} = {\sqrt{NT\tau}\mathcal{E}_\mathrm{eff}\eta}
\end{eqnarray}

Therefore, a reasonably large value of $\mathcal{E}_\mathrm{eff}$ improves the statistical sensitivity substantially. This quantity  must be calculated, and cannot be measured. However, the challenges in determining $\mathcal{E}_\mathrm{eff}$ arise purely due to relativistic effects~\cite{Sandars}. This necessitates a relativistic many-body treatment of this quantity. \\

Not all of the factors mentioned above can be satisfied by a single system. Several diatomic molecules have been proposed in the recent past, including HgX~\cite{HgX}, RaF~\cite{RaF}, PbF~\cite{PbF}, PbO~\cite{PbO}, and BaF~\cite{BaF,BaF1}, based on a combination of some of the considerations mentioned above. The search for new candidates that can promise better sensitivities than the current best leading molecules is crucial to future eEDM searches. \\

Polyatomic molecules are currently emerging as promising eEDM candidates. The first fast Sisyphus laser cooling of SrOH opened new avenues for polyatomics to come to the forefront of eEDM search experiments~\cite{SrOH}. Subsequently, RaOH was proposed as a suitable candidate~\cite{Isaev}. Around the same time, Kozyryev and Hutzler~\cite{Hutzler} proposed YbOH molecules for eEDM experiments. Unlike most diatomics, YbOH, a triatomic, offers two advantages at the same time, namely the possibility of laser-cooling, and possessing a bending mode with closely-spaced parity doublets, therefore having internal co-magnetometer states, and highly polarizable. Such a co-magnetometer state avoids systematics associated with reversing electric fields, while laser-cooling and trapping drastically reduces systematics such as a motional magnetic field. Trapping the molecules in an optical lattice would also offer a tremendous improvement in coherence time, of the order of seconds, as compared to the usual $\tau \sim$milliseconds. Lastly, the spectroscopy of the molecule has already been studied reasonably. In conclusion, based on these factors, the authors expect an increase of four orders in sensitivity as compared to the current best experiments. However, this would also require that $\mathcal{E}_\mathrm{eff}$ be comparable to that of leading eEDM candidates, and a calculation of the quantity has not been performed till date. In this work, we present accurate values of $\mathcal{E}_\mathrm{eff}$ and $d_M$ for the ground state of YbOH, in a linear geometry. We also make contact with experiment by presenting the first study of $\mathcal{E}_\mathrm{eff}$, for bent geometries. 

In this work, we present accurate values of $\mathcal{E}_\mathrm{eff}$ and $d_M$ for the ground state of YbOH, in a linear geometry. We also present our results for $\mathcal{E}_\mathrm{eff}$, for bent geometries that are relevant for eEDM search experiments. \\

The expectation value expressions for $\mathcal{E}_\mathrm{eff}$ and $d_M$ are, respectively \\

\begin{eqnarray}
\mathcal{E}_\mathrm{eff}&=& 2ic  \langle \psi \arrowvert \sum_{j=1}^{N_e} \beta \gamma_5 \, p_j^2 \arrowvert \psi \rangle, \\
d_M &=& \langle \psi \arrowvert (-\sum_{j=1}^{N_e} r_j + \sum_{A=1}^{N_A} Z_{A} r_A) \arrowvert \psi\rangle 
\end{eqnarray}

The summation, j, is over the number of electrons ($N_e$) in the molecule, while A denotes the summation over the number of nuclei ($N_A$). $\beta$ and refers to the Dirac beta matrix, and $\gamma_5$ is the product of the Dirac matrices. $p_j$ refers to the operator corresponding to the momentum of the $j^{th}$ electron. $r_j$ the position vector from the origin to the site of the $j^{th}$ electron. $Z_A$ is the atomic number of the $A^{th}$ nucleus, and $r_A$ is the vector from the origin to the $A^{th}$ nucleus. We work in the Born-Oppenheimer approximation, where the nuclei are `clamped', with respect to the electrons. In order to obtain $\mathcal{E}_\mathrm{eff}$ and $d_M$, we need to take the expectation values of the respective operators. Further details can be found in Ref.s~\cite{Abe,AEM,BPD}. \\

In order to evaluate the expressions given in eq.s (2) and (3), we use a relativistic coupled cluster method, where $\arrowvert \psi \rangle = e^T \arrowvert \Phi_0 \rangle$. Here, T is known as the cluster operator, and $\Phi_0$ is the reference state, which is the DF wave function in this case. Further details can be found in Ref.s ~\cite{Abe,HgX,AEM}. 
It is worth noting that at a given level of particle-hole excitation, the evaluation of the electron correlation effects by the coupled cluster method is equivalent to doing so to all orders in perturbation theory~\cite{Bishop}. Once we compute the wave function, the property of interest, O, can be obtained by evaluating an expectation value expression~\cite{Bartlett} \\

\begin{eqnarray}
\langle O \rangle &=& \frac{\langle \Phi_0 \arrowvert e^{T \dagger}Oe^T \arrowvert \Phi_0 \rangle}{\langle \Phi_0 \arrowvert e^{T \dagger}e^T \arrowvert \Phi_0 \rangle} \\
&=& \langle \Phi_0 \arrowvert O \arrowvert \Phi_0 \rangle \nonumber \\
&+& \langle \Phi_0 \arrowvert (1+T_1+T_2)^\dagger O_N (1+T_1+T_2) \arrowvert \Phi_0 \rangle_C 
\end{eqnarray}

The subscripts `N' and `C' mean that the creation and annihilation operators are normal ordered and that each term in the expression is fully connected, respectively~\cite{Kvas,Bishop,Lindgren}. We work with the coupled cluster singles and doubles (CCSD) approximation, where $T=T_1+T_2$. Also, although we solve the full CCSD equations to obtain the amplitudes, associated with the excitation operators $T_1$ and $T_2$, we only use the linear terms in T for solving the expectation value expression in Eq. (5). This approximation is reasonable, since dominant contributions to our properties of interest are mostly from the linear terms. \\

In Eq. (5), the first term is the DF expression. On expanding the second term, we get $\langle \Phi_0 \arrowvert O_N T_1 \arrowvert \Phi_0 \rangle_C$, $\langle \Phi_0 \arrowvert T_1^\dagger O_N T_1 \arrowvert \Phi_0 \rangle_C$, and so on. Hereafter, we shall write these terms in a more concise manner as $OT_1$, $T_1^\dagger O T_1$, etc. These terms correspond to different kinds of physical effects arising from electron correlation. 
One of the principal merits of the above approach is that it makes this connection transparent. \\

For our computations, we used the UTChem code~\cite{Utchem,utchem2} for DF and atomic orbital (AO) to molecular orbital (MO) integral transformations, while we obtained the t-amplitudes, from Dirac08~\cite{Dirac}. We performed our calculations for a linear geometry (the geometry of YbOH, as demonstrated in Ref.~\cite{JCP}), with the Yb-O bond length being 2.0026 Angstroms~\cite{ACV}, and O-H 0.922 Angstroms~\cite{ACV}. The DF computations require basis sets, which are single-particle functions, as an input~\cite{Dyall}. We employed gaussian type orbitals, specifically uncontracted Dyall's double zeta (DZ), triple zeta (TZ; which have more elements and are of better quality than the DZ option), and quadruple zeta (QZ; the largest that is available in Dyall's database) basis sets for all three atoms~\cite{Dyall2,Dyall3}. We round-off all of our calculated values in this work to two decimal places. \\

\begin{table}[h] 
\squeezetable
 \centering
 \begin{tabular}{ccccc}
 \hline
  Basis &$d_M^{DF}$&$d_M$&$\mathcal{E}_\mathrm{eff}^{DF}$&$\mathcal{E}_\mathrm{eff}$ \\
 \hline
 DZ &0.83&0.82&17.78&23.49\\
 TZ &0.90&1.02&18&23.85\\
 QZ &0.94&1.10&18.02&23.80\\
 \hline
 \end{tabular}
 \label{tab:contributions}
 \caption{The calculated values of $\mathcal{E}_\mathrm{eff}$ (in GV/cm), and $d_M$ (in Debye (D)). The Dirac-Fock (DF, in superscript), and the total (no superscript) contributions have been provided. The TZ result is at 800 au virtuals' cut-off value, and QZ at 500 au. }
\end{table}

Table I provides the results of our computations, for $\mathcal{E}_\mathrm{eff}$ and $d_M$. The results show that $\mathcal{E}_\mathrm{eff}$ is comparable to that of YbF. This is not too surprising, since in both systems, the dominant contributions come from the unpaired electron that belongs to the singly occupied MO of the Yb atom. However, we note that $d_M$ is substantially smaller than that for YbF (at 3.91 D~\cite{YbFPDM}). We did not impose any cut-off on the virtuals at the AO to MO transformation stage and therefore subsequently in the CCSD level, in the DZ calculations. However, due to the steep computational cost involved in the TZ and QZ calculations, we cut-off the high-lying virtuals, specifically those above 800 atomic units (au) for the TZ calculations. We chose this cut-off value, not only because it is sufficiently high-lying, but also because the energy difference between the cut-off value and the next virtual is 300 au. This choice strengthens our case, since the energy of the  next orbital is over 1000 au. For the QZ basis, we imposed a 200 au cut-off. From previous works, for example, Ref.~\cite{Sunaga2016}, we know that a cut-off of 200 au is sufficient to obtain accurate results. We still explicitly verify this by examining $\mathcal{E}_\mathrm{eff}$ and $d_M$ values for sample cut-off values using the QZ basis. We find that the CCSD value of $d_M$ hardly changes (1.10 D), while $\mathcal{E}_\mathrm{eff}$ changes by less than one percent (23.56, 23.72, and 23.80 for 100, 200, and 500 au respectively), which is well within the error bars in our calculations. \\


We now examine the correlation effects in $\mathcal{E}_\mathrm{eff}$, by examining the terms from eq. (5). The results (for the QZ basis) are presented in Tables II. We also give the QZ results for YbF~\cite{Abe}, for comparison. 
This illustrates the similarities not only in the DF values, but also in the correlation trends for the two isoelectronic molecules. These are interesting in their own right from a many-body theoretic point of view, and also from the perspective of what one may expect for the $\mathcal{E}_\mathrm{eff}$ of $YbCH_3$, which has also been proposed as an interesting eEDM candidate~\cite{Hutzler}. The similarity in the  $\mathcal{E}_\mathrm{eff}$ values of YbF, YbOH, and $YbOCH_3$ would not be a surprise though, since one would expect that the electronic structure of the system would be more important than the number of electrons themselves, particularly the number of electrons, and more importantly the fact that there is one unpaired electron in the same heavy atom, Yb. Since the expectation value approach and the energy derivative give almost the same result (23.1 and 23 GV/cm respectively~\cite{confproc}) for YbF, we expect that YbOH's $\mathcal{E}_\mathrm{eff}$ would not change if were to be calculated using an energy derivative. \\

\begin{table}[h!] 
 \centering
 \begin{tabular}{ccc}
 \hline
  Term&$\mathcal{E}_\mathrm{eff}$&$\mathcal{E}_\mathrm{eff}^{YbF}$ \\
 \hline
  DF &18.02&18.16\\
  $H_\mathrm{eEDM}^\mathrm{eff}T_1$ + cc &6.56&6.28\\
  $T^\dagger_1H_\mathrm{eEDM}^\mathrm{eff}T_1$&-0.86&-1.31\\
  $T^\dagger_1H_\mathrm{eEDM}^\mathrm{eff}T_2$ + cc&0.16&0.18\\
  $T^\dagger_2H_\mathrm{eEDM}^\mathrm{eff}T_2$ &-0.16&-0.17\\
  \hline
 \end{tabular}
 \caption{Contributions from the individual terms of the expectation value expression, at the QZ level of basis (with 200 au cut-off for YbOH), to $\mathcal{E}_\mathrm{eff}$ (GV/cm). cc refers to the complex conjugate of the term that it accompanies. }
\end{table}

The $d_M$s are very different for YbOH and YbF. In the case of YbOH, O can accept electrons from both Yb and H, while in the case of YbF, F can only do so from Yb. There is cancellation between the dipole moment contributions from `Yb to O' and `O to H', as they are in opposite directions. Therefore, $d_M$ of YbOH is smaller than that of YbF. This is qualitatively confirmed from the values of Mulliken charges of YbOH,  0.684 (Yb), -1.064 (O), and 0.380 (H), obtained at the DF level, using the DZ basis set (details of Mulliken analysis can be found in Ref.~\cite{Sun}). \\

We now present the error estimate in our calculations. The error in the calculated values can be due to exclusion of higher order correlation effects, and choice of basis sets. For $\mathcal{E}_\mathrm{eff}$, we look at the contributions from terms involving $T_2$ (which is negligibly small, since the contributions from  $T^\dagger_1H_{\mathrm{eEDM}}^{\mathrm{eff}}T_2$ and  $T^\dagger_2H_{\mathrm{eEDM}}^{\mathrm{eff}}T_2$ almost exactly cancel), and compare the sum with that obtained from terms involving $T_1$ (5.7 GV/cm). We expect that terms containing $T_3$ and beyond will contribute less than those that contain $T_2$, and therefore estimate the error from higher order excitations to be almost negligible, and conservatively set it at two percent. The finite field CCSD(T) results for $\mathcal{E}_\mathrm{eff}$ of several other molecules, from our previous work~\cite{FFCC}, indicate that both partial triples and the non-linear terms in the expectation value will not contribute significantly. Hence, a conservative error estimate by neglecting non-linear terms is about two percent. The error due to basis set incompleteness can be estimated as two percent, by assuming that the result would not change more than the difference between the TZ and QZ values. The combined error from these sources for $\mathcal{E}_\mathrm{eff}$ is about six percent. 
\\

We now turn to $\mathcal{E}_\mathrm{eff}$ at the DF level ($\mathcal{E}_\mathrm{eff}^{DF}$), for bent geometries, as shown in Figure I. Although the ground state of the YbOH molecule is linear, results for bent geometries are relevant for eEDM experiments that rely on low-lying vibrational states of the ground electronic states with parity-doubling~\cite{Hutzler}. In the case of YbOH, since the dominant contribution to $\mathcal{E}_\mathrm{eff}$ is at the DF level (about 75 percent), the results do not change significantly from DZ through QZ basis, and with the assumption that the correlation effects do not drastically change with bond angle, it suffices to compute the quantity at the DF level, with the DZ basis. \\

\begin{figure}[h!]
\centering
\includegraphics[scale=0.45]{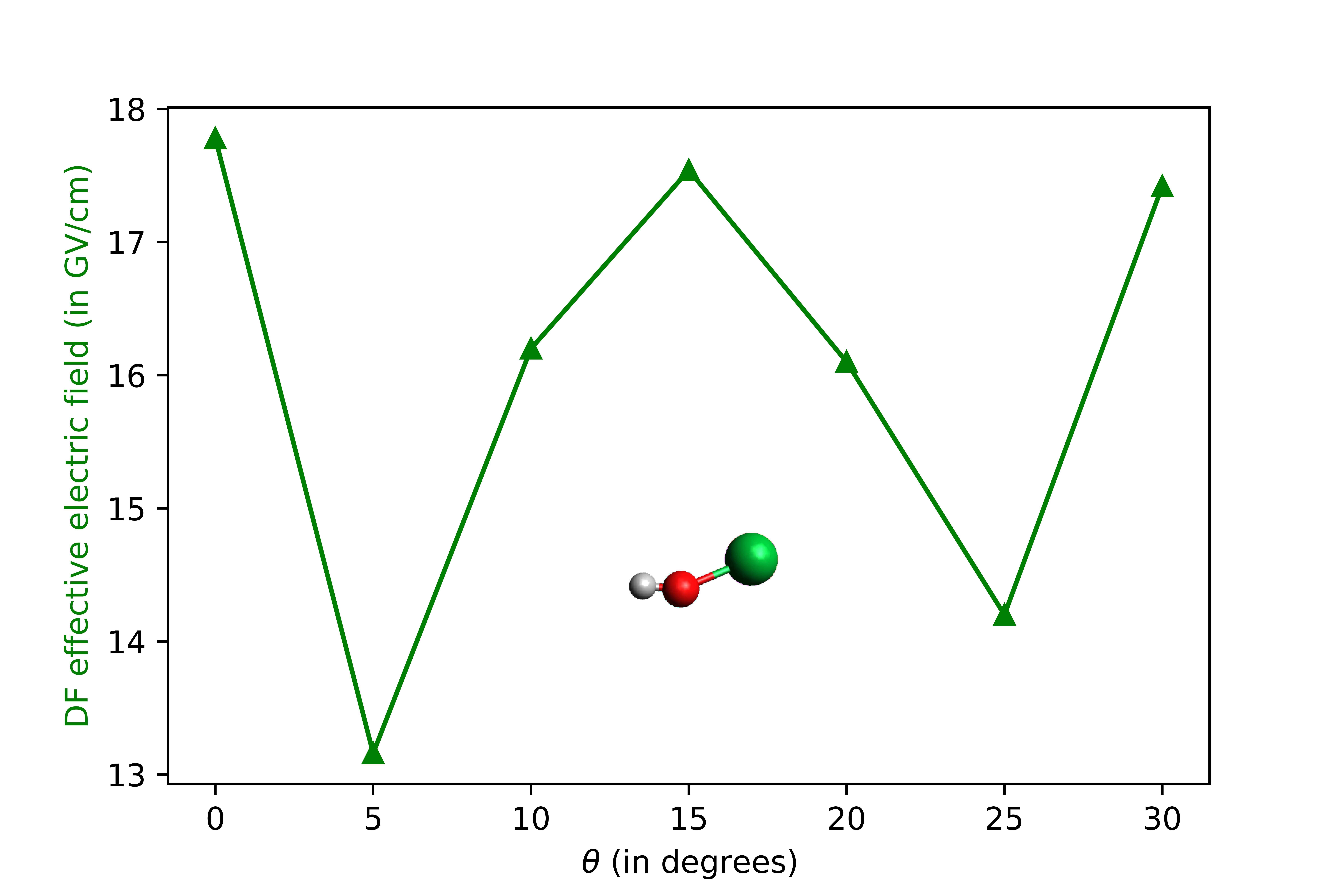}
\caption{The calculated values of $\mathcal{E}_\mathrm{eff}^{DF}$ (GV/cm) at different bond angles (in degrees). As the inset shows, in the H-O-Yb molecule, $\theta$ is the angle formed by O-Yb, with respect to H-O.}
\end{figure}

The Figure shows that $\mathcal{E}_\mathrm{eff}^{DF}$ varies from around 13 to 18 GV/cm. The final value of $\mathcal{E}_\mathrm{eff}$, after including the correlation effects, would be expected to change by about 24 percent, as in the linear case. We reiterate that the purpose of the analysis with the bent geometry is to demonstrate that $\mathcal{E}_\mathrm{eff}$ does not affect the prospects of a highly sensitive eEDM experiment, and to compute  $\mathcal{E}_\mathrm{eff}$ by capturing the most important physical effects. \\

We can rewrite the DF contribution as \\

\begin{eqnarray}
\mathcal{E}_\mathrm{eff}^{DF} = 2 \sum_{k,l} C_k^{L*} C_l^S \langle \chi_k^L \arrowvert h_{eEDM} \arrowvert \chi_l^S \rangle
\end{eqnarray}

where k and l are summations over the large and small component basis functions, themselves denoted by $\chi_k^L$ and $\chi_l^S$ respectively, Cs are the DF coefficients, and $h_{eEDM}$ is the one-body eEDM operator. We perform this analysis for three bond angles, in order to examine the dominant contribution in the equation given above, viz, $s-p_{1/2}$ and $p_{1/2}-s$ mixings of Yb, where the bra and ket in the above equation are s and $p_{1/2}$ for the former, and $p_{1/2}$ and s for the latter~\cite{HgXa}, for bent geometries. We obtained 13.07, 16.1, and 12.97 GV/cm for them for angles of 5, 10, and 15 degrees respectively, while the effective electric fields at the DF level for these angles are 13.16, 16.20,and 13.06 GV/cm respectively. The other mixings (such as $p_{3/2}$ and $d_{3/2}$ from Yb, as well as all mixings from the lighter atoms) contribute little to $\mathcal{E}_\mathrm{eff}^{DF}$, and we found them to usually be around 0.1 GV/cm or lesser, and therefore we have not presented those results here. The analysis shows that it is change in the terms involving mixings between s and $p_{1/2}$ that is responsible for the change in $\mathcal{E}_\mathrm{eff}^{DF}$. \\


In Table III we discuss the variation of $\mathcal{E}_\mathrm{eff}^{DF}$ with bond length ($R_{Yb-O}$), around the chosen value. In a vibrational state (which is the state of interest in the polyEDM experiment), specifically in a stretch mode, the bond length changes. Therefore, an estimate of how $\mathcal{E}_\mathrm{eff}$ changes with bond length becomes relevant. We observe that $\mathcal{E}_\mathrm{eff}^{DF}$ increases with $R_{Yb-O}$. However, we also note that $\mathcal{E}_\mathrm{eff}^{DF}$ tends towards saturation, as $R_{Yb-O}$ increases. \\

\begin{table}[h] 
 \centering
 \begin{tabular}{cc}
 \hline
 $R_{Yb-O}$&$\mathcal{E}_\mathrm{eff}^{DF}$\\
 \hline
 1.5026&12.24\\
 1.7526&15.78\\
 2.0026&17.78\\
 2.2526&18.73\\
 2.5026&18.97\\
 \hline
 \end{tabular}
 \label{tab:contributions}
 \caption{The calculated values of $\mathcal{E}_\mathrm{eff}^{DF}$ (in GV/cm) at different bond lengths, $R_{Yb-O}$ (in Angstroms). A bond length of 2.0026 Angstrom corresponds to the original value. }
\end{table}

We do not expect $\mathcal{E}_\mathrm{eff}^{DF}$ to change significantly with $R_{O-H}$. However, for the sake of completeness, we present the results in Table IV. \\

\begin{table}[h] 
 \centering
 \begin{tabular}{cc}
 \hline
 $R_{O-H}$&$\mathcal{E}_\mathrm{eff}^{DF}$\\
 \hline
 -0.422&18.14\\
 -0.922&17.78\\
 -1.422&17.49\\
 \hline
 \end{tabular}
 \label{tab:contributions}
 \caption{The calculated values of $\mathcal{E}_\mathrm{eff}^{DF}$ (in GV/cm) at different bond lengths (absolute values), $R_{O-H}$ (in Angstroms). We had chosen the O atom as our origin. The original value of bond length is 0.922 Angstroms.  }
\end{table}

Finally, in Table V, we present $\mathcal{E}_\mathrm{eff}^{DF}$, with both the lengths varied. The trend remains, with $\mathcal{E}_\mathrm{eff}^{DF}$ increasing with the bond lengths, and reaching towards a `saturation' point. We expect that the inclusion of correlation effects will not change the trends, but only shift each of the values obtained at the DF level. \\

\begin{table}[h] 
 \centering
 \begin{tabular}{ccc}
 \hline
 $R_{Yb-O}$&$R_{O-H}$&$\mathcal{E}_\mathrm{eff}^{DF}$\\
 \hline
 1.5026&0.422&12.17\\
 1.7526&0.672&15.87\\
 2.0026&0.922&17.78\\
 2.2526&1.172&18.58\\
 2.5026&1.422&18.58\\
 \hline
 \end{tabular}
 \label{tab:contributions}
 \caption{The calculated values of $\mathcal{E}_\mathrm{eff}$ (in GV/cm) at different bond lengths (absolute values), $R_{Yb-O}$ (in Angstroms). We had chosen the O atom as our origin. The combination of 2.0026 and 0.922 Angstroms corresponds to the original value. }
\end{table}

In conclusion, we have calculated $\mathcal{E}_\mathrm{eff}$ and $d_M$ of the YbOH molecule in its ground state, using a relativistic coupled cluster method. The results show that $\mathcal{E}_\mathrm{eff}$ is almost as large as that in YbF, at 23.72 GV/cm. We also examine $\mathcal{E}_\mathrm{eff}$ and $d_M$ at different cut-off values for the virtual orbitals. We estimate the errors in our calculations of $\mathcal{E}_\mathrm{eff}$ to be about six percent, due to various sources like basis set incompleteness, exclusion of higher order excitations, and ignoring the terms that are non-linear in the cluster operator in the expectation value. We also present relativistic mean field values for a bent geometry. We tried to understand the dependence of the $\mathcal{E}_\mathrm{eff}^{DF}$ on bond angle, by studying the mixings of orbitals in YbOH. Our analysis shows that the mixing between the heavier atom's s and $p_{1/2}$ orbitals is dominant, and its variation is responsible for the value of $\mathcal{E}_\mathrm{eff}^{DF}$ changing with angle. These calculations will provide useful inputs for the feasibility of an eEDM experiment with YbOH. \\ 

\section*{Acknowledgements}
The computations were performed on VIKRAM-100 cluster, Physical Research Laboratory, Ahmedabad, India, as well as on the CHIYO work station, Tokyo Institute of Technology, Tokyo, Japan. We acknowledge Amar Vutha for  thought-provoking discussions and useful suggestions, and Prof. T. Steimle for communication on the bond lengths.

\end{document}